\documentclass{aa}  
\usepackage{ulem}
\usepackage{color}
\usepackage{graphicx}
\usepackage[cmyk]{xcolor}
\usepackage{txfonts}
\usepackage[colorlinks,
            linkcolor=blue,
            anchorcolor=blue,
            citecolor=blue,
            urlcolor=blue]{hyperref}
\usepackage{amsmath}
\usepackage{amssymb}
\usepackage{color}
\usepackage{makecell}
\usepackage{array}
\usepackage{caption}
\usepackage[ruled,vlined]{algorithm2e}
\usepackage{float}
\begin{document}

   \title{Classification of real and bogus transients using active learning and semi-supervised learning}

   \subtitle{}

   \author{Yating Liu
          \inst{1, 6}\thanks{Email: liuyat@mail.ustc.edu.cn}
          \and
          Lulu Fan
          \inst{2, 3, 4}\thanks{Email: llfan@ustc.edu.cn}
          \and
          Lei Hu
          \inst{5, 7}
          \and
          Junqiang Lu
          \inst{2, 3}
          \and
          Yan Lu
          \inst{6}
          \and
          Zelin Xu
          \inst{2, 3}
          \and
          Jiazheng Zhu
          \inst{2, 3}
          \and
          Haochen Wang
          \inst{2, 3}
          \and
          Xu Kong
          \inst{2, 3, 4}
          }

   \institute{School of Artificial Intelligence and Data Science, University of Science and Technology of China, Hefei, 230026, China
        \and
             CAS Key Laboratory for Research in Galaxies and Cosmology, Department of Astronomy, University of Science and Technology of China, Hefei, 230026, China
        \and
            School of Astronomy and Space Science, University of Science and Technology of China, Hefei, 230026, China
        \and
            Deep Space Exploration Laboratory, Hefei, 230088, China
        \and
            McWilliams Center for Cosmology, Department of Physics, Carnegie Mellon University, 5000 Forbes Ave, Pittsburgh, 15213, PA, USA
        \and
             Shanghai AI Laboratory, Shanghai, 200232, China
        \and
             Purple Mountain Observatory, Nanjing, 210023, China 
             }
   \date{Accepted December 1, 2024}

 
  \abstract
   {The mounting data stream of large time-domain surveys renders the visual inspections of a huge set of transient candidates impractical. Techniques based on deep learning-based are popular solutions for minimizing human intervention in the time domain community. The classification of real and bogus transients is a fundamental component in real-time data processing systems and is critical to enabling rapid follow-up observations. Most existing methods (supervised learning) require sufficiently large training samples with corresponding labels, which involve costly human labeling and are challenging in the early stages of a time-domain survey. One method that can make use of training samples with access to only a limited amount of labels is highly desirable for future large time-domain surveys. These include the forthcoming 2.5-meter Wide-Field Survey Telescope (WFST) six-year survey and the Vera C. Rubin Observatory Legacy Survey of Space and Time (LSST).}
   {Deep-learning-based methods have been favored in astrophysics owing to their adaptability and remarkable performance. They have been applied to the task of the classification of real and bogus transients. Unlike most existing approaches, which necessitate massive and expensive annotated data, we aim to leverage training samples with only 1000 labels and discover real sources that vary in brightness over time in the early stages of the WFST six-year survey.}
   {We present a novel deep learning method that combines active learning and semi-supervised learning to construct a competitive real-bogus classifier. Our method incorporates an active learning stage, where we actively select the most informative or uncertain samples for annotation. This stage aims to achieve higher model performance by leveraging fewer labeled samples, thus reducing annotation costs and improving the overall learning process efficiency. Furthermore, our approach involves a semi-supervised learning stage that exploits the unlabeled data to enhance the model's performance and achieve superior results, compared to using only the limited labeled data.}
   {Our proposed methodology capitalizes on the potential of active learning and semi-supervised learning. To demonstrate the efficacy of our approach, we constructed three newly compiled datasets from the Zwicky Transient Facility (ZTF), achieving average accuracies of 98.8\%, 98.8\%, and 98.6\% across these three datasets. It is important to note that our newly compiled datasets only work in terms of testing our deep learning methodology and there may be a potential bias between our datasets and the complete data stream. Therefore, the observed performance on these datasets cannot be assumed to directly translate to the general alert stream for general transient detection in actual scenarios. The algorithm will be integrated into the WFST pipeline, enabling an efficient and effective classification of transients in the early period of a time-domain survey.}
   {}


   \maketitle
%

\section{Introduction} \label{sec:intro}

\begin{figure*}[t]
\centering
\includegraphics[width=0.9\linewidth]{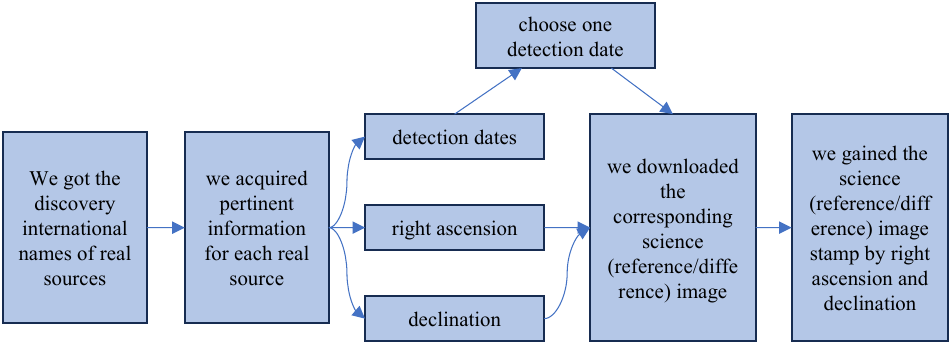} 
\caption{Process of collecting real sources. The outputs of this process are real sources and each real detection includes a 63 $\times$ 63 science image stamp, the corresponding 63 $\times$ 63 reference image stamp, and the corresponding 63 $\times$ 63 difference image stamp.
\label{fig:real_source}}
\end{figure*}

\begin{figure*}[t]
\centering
\includegraphics[width=1\linewidth]{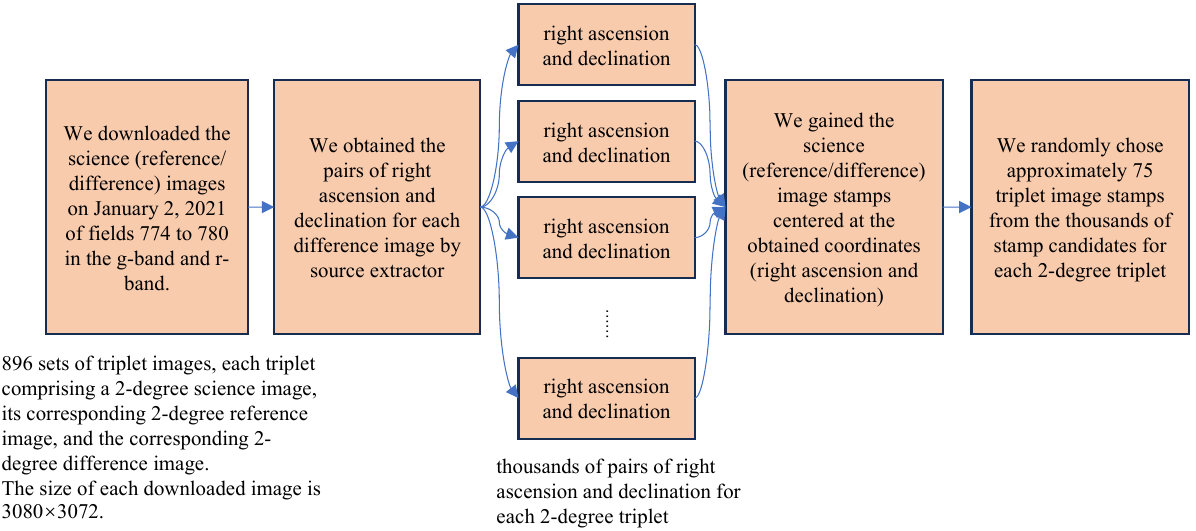} 
\caption{Pipeline of gathering bogus sources. The outputs of this pipeline are bogus detections and each bogus detection includes a 63 $\times$ 63 science image stamp, the corresponding 63 $\times$ 63 reference image stamp, and the corresponding 63 $\times$ 63 difference image stamp.
\label{fig:bogus_detection}}
\end{figure*}

\begin{figure*}[t]
\centering
\includegraphics[width=0.75\linewidth]{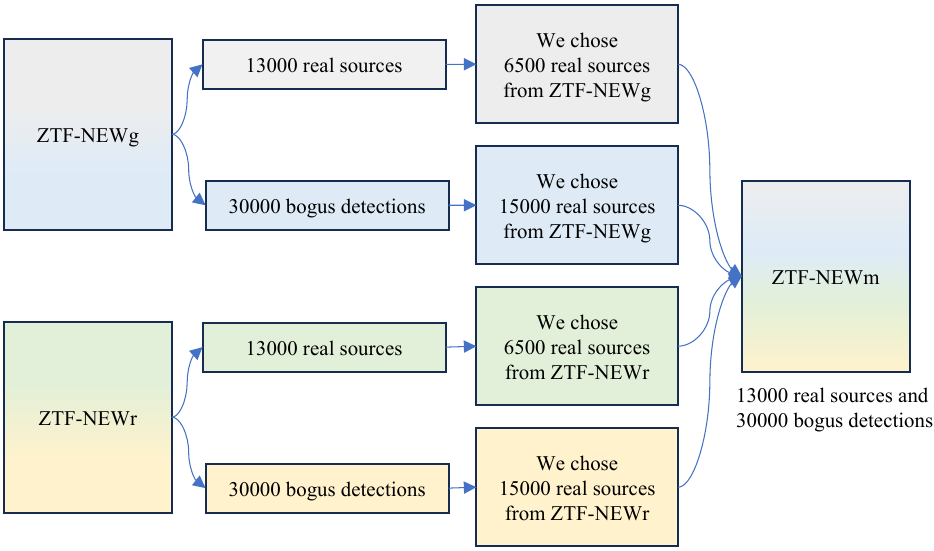} 
\caption{Construction pipeline of the ZTF-NEWm dataset. Both the ZTF-NEWg dataset and the ZTF-NEWr dataset contain 13000 real sources and 30000 bogus sources. We selected 6500 real sources from the ZTF-NEWg dataset and 6500 real detections from the ZTF-NEWr dataset. The 13000 selected samples were combined to form the real source component of the ZTF-NEWm dataset and above them contained at least 4000 first detection sources. Similarly, we randomly chose 15000 bogus sources from the ZTF-NEWg dataset and 15000 bogus detections from the ZTF-NEWr dataset. These chosen samples were then combined to create the bogus detection component of the ZTF-NEWm dataset. By combining the real source and bogus detection components, we obtained a dataset (ZTF-NEWm) consisting of 13000 real sources and 30000 bogus detections.
\label{fig:ZTF-NEWm}}
\end{figure*}

\begin{figure*}[t]
\centering
\includegraphics[width=0.84\linewidth]{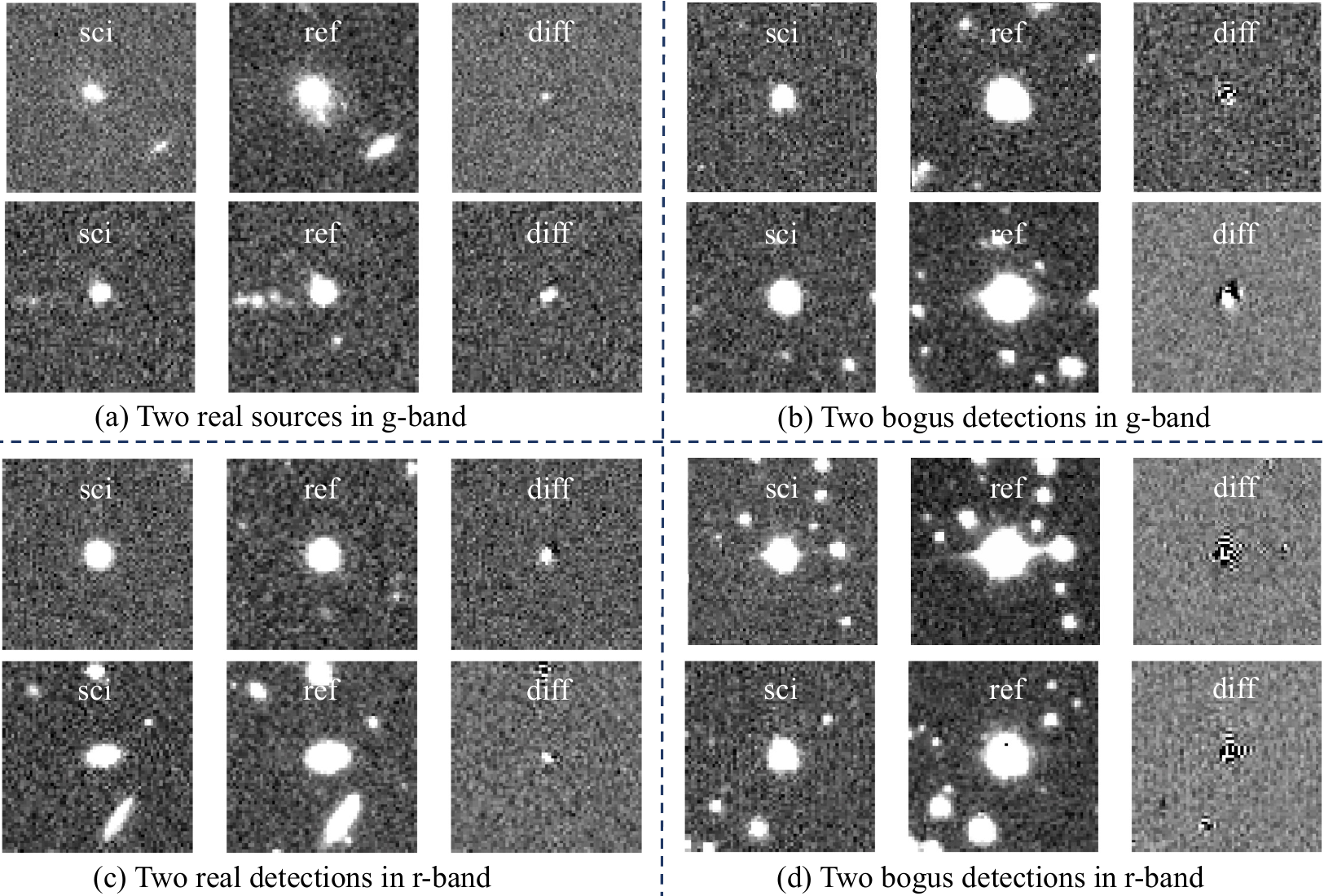} 
\caption{Examples of 63 pixels $\times$ 63 pixels triplet image stamps of our compiled datasets. The top-left part of the figure depicts two real sources in g-band and each row represents a real detection, showing the science image stamp, the reference image stamp, and the difference image stamp of the real detection from left to right respectively. Similarly, the bottom-left,  top-right, and bottom-right of this figure display two real detections in the r-band, two bogus detections in the g-band, and two bogus detections in the r-band respectively. 
\label{fig:example_real_bogus}}
\end{figure*}

\begin{figure*}[t]
\centering
\includegraphics[width=\linewidth]{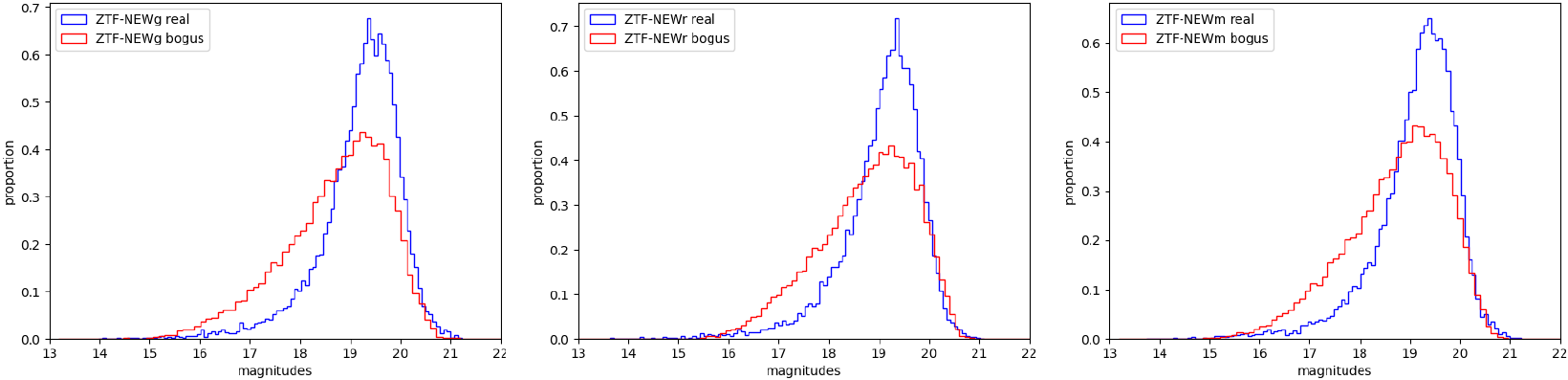} 
\caption{Distribution of magnitudes for real and bogus sources of the three datasets.
\label{fig:mag}}
\end{figure*}

Large-scale surveys are generating considerable data that record abundant
dynamic events in the sky, as exemplified by  Panoramic Survey Telescope and Rapid Response System (Pan-STARRS; \citealt{chambers2016pan}),  High Cadence Transient Survey (HiTS; \citealt{forster2016high}),  Dark Energy Survey (DES; \citealt{dark2005dark}),  Zwicky Transient Facility (ZTF; \citealt{bellm2018zwicky}),  forthcoming 2.5-meter Wide-Field Survey Telescope (WFST; \citealt{wang2023wfst}) six-year survey, and  Vera C. Rubin Observatory Legacy Survey of Space and Time (LSST; \citealt{abell2009lsst}).
Real detections that interest astronomers are events that vary in brightness over time, including variable, transient, and moving sources. It is critical to discover these real sources to enable rapid follow-up scientific observations (e.g., \citealt{hu2022prospects}).

In most surveys, three images are utilized to discover real detections (\citealt{ayyar2022identifying, duev2019real, reyes2023multi}):
(1) a reference image, which is an archival observation of the sky; (2) a science image, which is a recent image of the same sky area and the same band as the reference image; and 
(3) a difference image, which is the result of subtracting the reference image from the science image.
In principle, real detections can be discovered by the difference image with brightness residual (\citealt{hosenie2021meercrab}). However, most of the detection candidates detected by brightness residual are bogus detections caused by poor image subtraction, atmospheric dispersion, optical ghosting, etc. To filter out these bogus sources, an effective real-bogus classifier is essential for large sky surveys.

The mounting data stream of large time-domain surveys renders astronomers unable to visually inspect source candidates manually (\citealt{killestein2021transient, makhlouf2022train}), requiring new tools to process the overwhelming data (\citealt{Bailey2007How, Brink2013Using, Goldstein2015Automated, Sedaghat2017Effective, wright2015Machine}).
Deep-learning-based techniques have led to great progress in computer vision (\citealt{he2017mask, ren2015faster}) and serve as popular solutions for minimizing human intervention in the time-domain community (\citealt{duev2019real, Gomez2020Classifying, Turpin2020Vetting}). 
There are some methods \citep{cavanagh2021morphological, goode2022machine, reyes2018enhanced} using deep learning to tackle this real-bogus classification task to detect real sources. Several approaches have constructed their neural network architectures such as BRAAI \citep{duev2019real}, MeerCRAB \citep{hosenie2021meercrab}, and Deep-HiTS \citep{cabrera2017deep}. Moreover, some works have presented integrated systems, such as combining classifications from volunteers participating in the science project with those from a convolutional neural network \citep{wright2017transient}, including majority voting of boosting, random forest, and deep neural networks \citep{morii2016machine}.
The aforementioned works in the real-bogus classification task displayed a commendable performance, but they are still supervised learning approaches built on sufficiently large training samples with corresponding labels. This means they require costly human annotation and are challenging in the early period of a time-domain survey.
Furthermore, the model trained on data from another telescope should not be used directly, because data from different telescopes violate independent and identically distributed (i.i.d) which limits the consistency and high performance of classification (\citealt{hosenie2021meercrab}).
Therefore, a method that can utilize training samples with only a few  available labels is highly desirable for future large time-domain surveys such as the WFST and the LSST.

In this paper, we propose RB-C1000, a novel deep learning method using active learning and semi-supervised learning that obtains competitive real-bogus classifier results with only 1000 labels. 
We also have constructed three new datasets from the ZTF to demonstrate the effectiveness of our approach. It should be acknowledged that the three datasets only work in testing our deep learning method. Therefore, the observed performance on these datasets may not accurately represent real world scenarios.
The remaining parts of our paper are organized as follows. In Section \ref{sec:data}, we describe the construction pipeline of newly compiled datasets. Afterward, we provide a brief introduction to active learning and semi-supervised learning and we present the architecture of our novel deep learning method RB-C1000 in Section \ref{sec:method}. The experimental results are provided in Section \ref{sec:experiments}. Finally, we summarize our work in the last section.

\section{Data} \label{sec:data}

\begin{figure*}[t]
\centering
\includegraphics[width=1.0\linewidth]{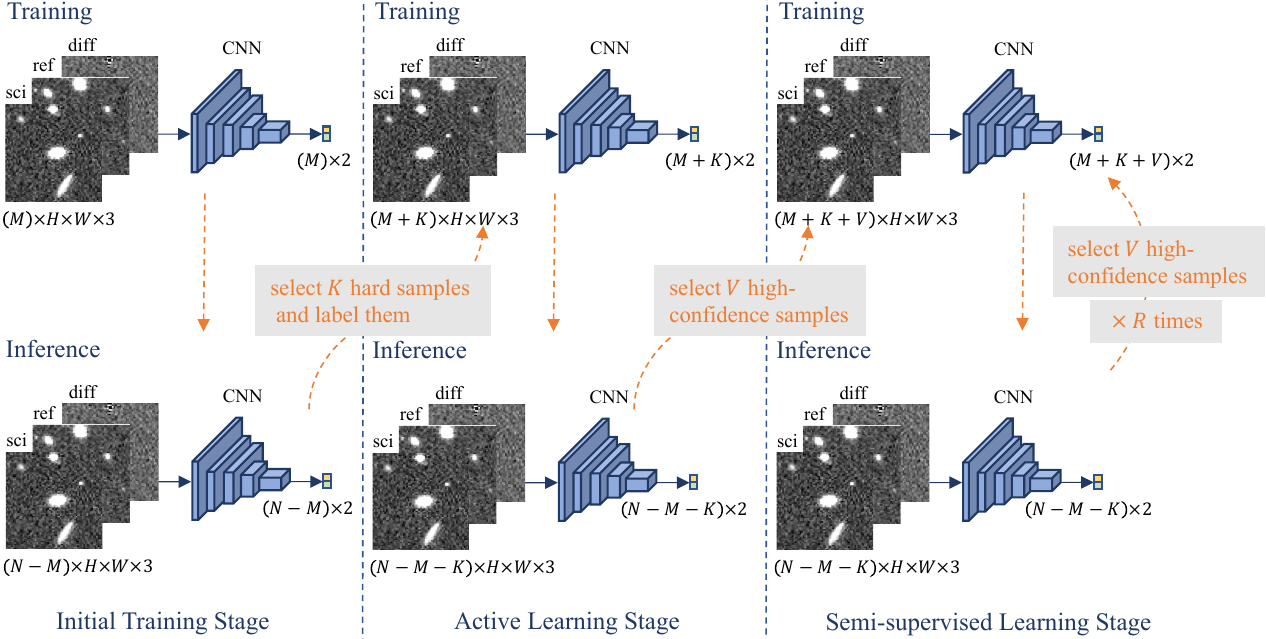} 
\caption{Architecture of our method. In the ITS, each labeled sample undergoes convolutional neural network processing to train an initial model. Subsequently, domain experts annotate the $K$ most challenging samples, as determined by the initial model's judgments. During the ALS, we employ the combined set of $(M+K)$ labeled samples to train an active training model. From this model, we select the top $V$ samples with high-confidence predictions and assign pseudo-labels to them. In the SSLS, we utilize the expanded dataset of $(M+K+V)$ samples to train a semi-supervised training model. This process is repeated for a total of $R$ iterations to obtain the final results.
\label{fig:pipeline}}
\end{figure*}

\begin{figure*}[t]
\centering
\includegraphics[width=1.0\linewidth]{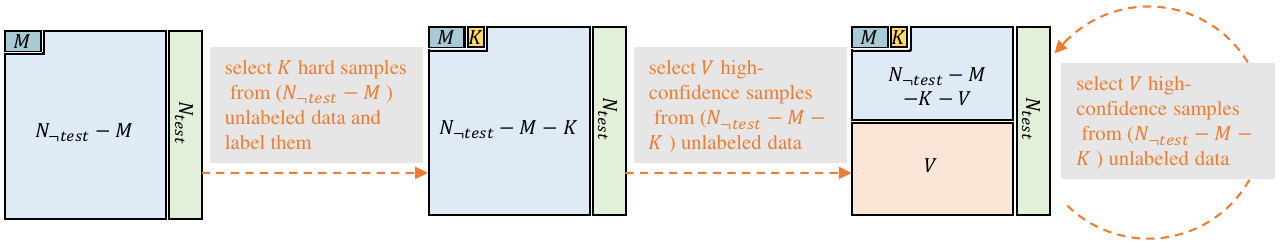} 
\caption{Process of labeling or pseudo-labeling for unlabeled samples. Following the training phase in the ITS, we select $K$ hard samples from the pool of  $(N_{\neg test}-M)$ unlabeled data and label them with expert annotations. Subsequently, after the training phase in the ALS, we choose $V$ over-threshold samples from $(N_{\neg test}-M-K)$ unlabeled data to assign pseudo-labels. Moving forward, after the training process in SSLS in the $r$-th iteration, we again select $V$ samples from the pool of $(N_{\neg test}-M-K)$ unlabeled data that exceed the predetermined threshold and assign pseudo labels. In the $(r+1)$-th iteration, we retrain the semi-supervised training model from scratch, refining the performance of our models in subsequent iterations.
\label{fig:sample}}
\end{figure*}

We have constructed new real-bogus classification datasets from the Zwicky Transient Facility (ZTF) to verify the effectiveness of our approach. To explore the effect of bands on the real-bogus results, we collected two single-band datasets (ZTF-NEWg: g-band, ZTF-NEWr: r-band) and one mixed-band dataset (ZTF-NEWm: taking half of the g-band data and half of the r-band data). The data collection process is divided into real source collection and bogus detection collection. The construction pipeline of newly compiled datasets is presented below.

Real sources:
To obtain real detections, we curated the discovery international names \footnote{\url{https://www.wis-tns.org/}} of real sources discovered by the ZTF from March 1, 2019 to March 1, 2022, such as ZTF20acmvsdj, ZTF21acjanhz. Through the collected names, we acquired pertinent information \footnote{\url{https://alerce.online/}} for each real source, including detection dates (observation UT date and time), right ascension, and declination. Generally, each real detection is observed and detected by ZTF multiple times during its lifetime. To prevent information leaks during the data partitioning process for training and evaluation, we chose one of the detection dates for each real source (\citealt{reyes2023multi}). Subsequently, we downloaded the corresponding science (reference and difference) image \footnote{\url{https://irsa.ipac.caltech.edu/applications/ztf/?__action=layout.showDropDown&view=Search}} of the specific detection date we chose and gained the science (reference and difference) image stamp (\citealt{Bertin2005SWarp}) by the right ascension and the declination. The process of collecting real sources is shown in Figure \ref{fig:real_source}. Finally, we obtained a total of 13055 g-band real sources and 13006 r-band real sources. We randomly removed 55 sources from the 13055 g-band real sources to obtain the exact 13000 g-band real detections as the real part of the ZTF-NEWg dataset. Similarly, we randomly deleted six detections from 13006 r-band real sources as the real part of the ZTF-NEWr dataset. To increase the challenge of our datasets, out of 13000 real sources collected in the g-band, 4000 sources were dated with their first detection dates, as sources from the initial detection date are typically fainter. The remaining 9000 sources were dated with their randomly sampled detection dates. The collection of real detections in the r-band followed the same pattern.

Bogus detections:
Initially, we downloaded the science (reference and difference) images on a specific day of specific fields in the g-band and r-band. We chose fields 774 to 780 on January 2, 2021 for the purposes of this paper. Each field contains 64 triplet images per band, with each triplet comprising a two-degree science image, its corresponding two-degree reference image, and the corresponding two-degree difference image, resulting in 896 sets of triplet images. Each downloaded image had dimensions of 3080 pixels $\times$ 3072 pixels. Subsequently, we obtained the pairs of right ascension and declination for detection candidates in each difference image by source extractor \citep{sex}. We gained the science (reference and difference) image stamps centered at the obtained coordinates and the size of each science (reference and difference) image stamp is 63 pixels $\times$ 63 pixels. To ensure the diversity of bogus sources, we conducted a random selection process. From the thousands of stamp candidates available for each two-degree triplet, we randomly chose approximately 75 triplet image stamps. The ratio of bogus detections to real detections in each two-degree triplet is overwhelmingly skewed, with bogus detections comprising approximately 99 percent of the total \citep{killestein2021transient}. Given this significant disparity, the chosen sampling approach is justified. The pipeline of collecting bogus detections is shown in Figure \ref{fig:bogus_detection}. 

After completing the above steps, we obtained a g-band dataset (ZTF-NEWg) and a r-band dataset (ZTF-NEWr) respectively. The mixed-band dataset selected half of the samples from the ZTF-NEWg and the ZTF-NEWr (shown in Figure \ref{fig:ZTF-NEWm}). The three newly compiled datasets for the real-bogus classification task, each with 13000 real sources and 30000 bogus detections, as shown in Figure \ref{fig:example_real_bogus}. The top left part of the figure depicts two real sources in g-band and each row represents a real detection, showing the science image stamp, the reference image stamp, and the difference image stamp of the real detection from left to right respectively. Similarly, the bottom left, the top right, and the bottom right of this figure display two real detections in the r-band, two bogus detections in the g-band, and two bogus detections in the r-band, respectively. The distribution of magnitudes for real and bogus sources of the three datasets are shown in Figure \ref{fig:mag}.

It should be noted that there is a potential bias between our datasets and the complete data stream, which exists in both real and bogus sources. Regarding real sources, the kinds of real detection are not diverse. They only consist of supernovas (SNe) as reported to the Transient Name Server (TNS) and lack large stellar flares, AGN variability, SNe below the nominal detection limit, and so on. For bogus detections, they might contain a few real detections, since they are randomly drawn from detection candidates. Our newly compiled datasets only work in terms of testing our deep-learning algorithm, as well as the effectiveness of each component in our method. Therefore, the performance of the three datasets cannot be assumed to apply to the general alert stream for detecting general transients in actual sky surveys.

\section{Method} \label{sec:method}

\subsection{Active learning and semi-supervised learning} \label{subsec:active}

Active learning:
The purpose of active learning \citep{prince2004does, settles2009active} is to use as few sample annotations as possible to achieve the best performance of the model. In many cases, it is impractical to obtain a large number of labeled samples due to limited resources. Active learning addresses this challenge by selecting the most informative samples for annotation to maximize the performance of the model. Specifically, there are a few labeled samples in the whole dataset, $M \ll N$, where $M$ and $N$ represent the number of labeled and total samples in the dataset, respectively. The active learning process begins by training an initial model using all available labeled samples. Then experts annotate the $K$ most challenging samples from the unlabeled data, as determined by the initial model's predictions. Finally, we leverage the $(M+K)$ labeled samples to train a supervised or semi-supervised model. By selectively annotating the most informative samples, active learning integrates artificial experience and additional expert knowledge into the model, leading to improving the model's ability to make accurate predictions.

Semi-supervised learning: 
Semi-supervised learning \citep{zhu2005semi, chapelle2009semi} has received increasing attention in recent years because human annotation is costly. The goal of semi-supervised methods is to enhance experimental outcomes by using unlabeled samples, as these data provide valuable distributional information. To be specific, regardless of whether the samples are labeled or not, they all originate from the same distribution. Numerous classic approaches have emerged in the field of semi-supervised learning such as semi-supervised support vector machines \citep{bennett1998semi, joachims1999transductive} and graph-based semi-supervised learning \citep{yang2016revisiting}. In this article, we employ the technique of pseudo-labeling. Concretely, after training a model by $(M+K)$ labeled samples, the model assigns a pseudo-label and a corresponding confidence score to each remaining sample in the dataset. And then a new model is retrained using both the labeled samples and the unlabeled data whose confidence scores exceed a predetermined threshold. By leveraging the power of unlabeled data, the retrained model can effectively harness the valuable information present in the unlabeled samples, leading to enhanced performance without incurring additional labor costs.

\subsection{RB-C1000 architecture} \label{subsec:architecture}
The pipeline of our approach is shown in Figure \ref{fig:pipeline}. 
Distinguishing itself from other methods, our approach is tailored to scenarios, where only a limited number of labeled samples are available ($M \ll N$, where $M$ and $N$ represent the number of labeled and total samples in the dataset respectively). Inspired by active learning and semi-supervised learning, our novel deep-learning method, RB-C1000, consists of three components: the initial training stage (ITS), the active learning stage (ALS), and the semi-supervised learning stage (SSLS). To minimize manual annotation, the ITS utilizes the idea of active learning which aims to label the $K$ hardest samples from the pool of unlabeled data as determined by the initial model's predictions. The objective of the ALS is to search for the $V$ most confident samples from unlabeled data, as assessed by the active training model, to aid the training process in the SSLS. During the SSLS, we iterate the training of a semi-supervised model multiple times to obtain promising and accurate results. Each module works together to achieve performance improvements and further details of each stage are elaborated as follows:

\begin{algorithm}[t]
\caption{Framework of RB-C1000.}
\label{alg:algorithm1}
\tcp{\color{blue}****************************************}
\tcp{\color{blue}The initial training stage}
\tcp{\color{blue}****************************************}
\KwIn{Initial supervised model: $f_{init}$; Data: total $N$ samples, $M$ labeled samples, $N_{test}$ test samples, $(N_{\neg test}-M)$ unlabeled samples, the count of challenging samples $K$}
Training the initial supervised model $f_{init}$ with {\color{red}$M$ labeled data}\\
\KwOut{The $K$ hard samples from $(N_{\neg test}-M)$ unlabeled data judged based on $f_{init}$; The results of $N_{test}$ test samples}  
        
\tcp{\color{blue}****************************************}      
\tcp{\color{blue}The active learning stage}
\tcp{\color{blue}****************************************}    
\KwIn{Active training model: $f_{act}$; predetermined threshold $\tau$; Data: total $N$ samples, $M$ labeled samples, $N_{test}$ test samples, {\color{red}the $K$ hard samples annotated by experts}, $(N_{\neg test}-M-K)$ unlabeled samples}
1. Training the active training model $f_{act}$ with {\color{red}$(M + K)$ labeled data}\\
2. Selecting the $V$ samples from $(N_{\neg test}-M-K)$ unlabeled data with confidence scores surpassing a predetermined threshold $\tau$ based on $f_{act}$ and assign pseudo-labels to them\\
\KwOut{The $V$ highest confidence samples and their corresponding pseudo labels; The results of $N_{test}$ test samples}
        
\tcp{\color{blue}****************************************}        
\tcp{\color{blue}The semi-supervised learning stage}
\tcp{\color{blue}****************************************}
\KwIn{Semi-supervised training model: $f_{semi}$; predetermined threshold $\tau$;         Data: total $N$ samples, $(M + K)$ labeled samples, $N_{test}$ test samples, {\color{red}The $V$ highest confidence samples and their corresponding pseudo labels}, $(N_{\neg test}-M-K)$ unlabeled samples}
\For{$r=1; r\leq R; r++$}{
1. Training the semi-supervised training training model $f_{semi}^r$ with {\color{red}$(M + K + V)$ samples}\\
2. Searching the $V$ samples from $(N_{\neg test}-M-K)$ unlabeled data with confidence scores surpassing a predetermined threshold $\tau$ judged based on $f_{semi}^r$}
\KwOut{The results of $N_{test}$ test samples}         
\end{algorithm}

\begin{table*}[t]
\renewcommand\arraystretch{1.3}
\begin{center}
\caption{A summary of the basic parameters in Section \ref{subsec:setting}.}\label{ex_p}
\begin{tabular}{c|c|c} 
\hline
Parameters & Corresponding meanings & Corresponding assignments \\
\hline
\multicolumn{3}{c}{Data}\\
\hline
$N$ & the number of total samples of each dataset & 43000\\
$N_{test}$ & the count of test samples of each dataset & 4300\\
$M$ & the number of labeled samples of each dataset & 900\\
$K$ & the count of challenging samples judged by the initial model & 100\\
$H$ & the height of a image stamp & 63 pixels\\
$W$ & the width of a image stamp & 63 pixels\\
\hline
\multicolumn{3}{c}{Model}\\
\hline
random horizontal flipping & data augmentation & 0.5\\
random vertical flipping & data augmentation & 0.5\\
batch size & the number of samples taken by an iteration & 256\\
epoch & the number of times for training the entire dataset & 100\\
$\tau$ & predetermined threshold & 0.95\\
$R$ & the number of repetitions of the SSLS phase & 3\\
learning rate & \makecell[c]{the step size of the gradient moving toward the optimal \\solution of the loss function in each iteration} & 2e-4\\
\hline
\end{tabular}
\end{center}
\end{table*}

Initial training stage: In the ITS,  every labeled sample is passed through a convolutional neural network (CNN; \citealt{lecun1989backpropagation}) to train an initial supervised model. Following the training process, the remaining $(N-M)$ samples are sent into the initial model, producing $(N-M)$ 2-dimensional results, as the left of Figure \ref{fig:pipeline} shows. The first and the second dimensions of each result represent the probability of bogus detection and real detection respectively, with their sum being equal to 1.
Subsequently, we collected the maximum value from each 2-dimensional result of the $(N_{\neg test}-M)$ unlabeled samples and get $(N_{\neg test}-M)$ one-dimensional (1D) results. These $(N_{\neg test}-M)$ 1-dimensional results represent the confidence of the initial model's prediction and we arrange these results in ascending order. Here, $N_{\neg test}$ denotes the count of the dataset excluding test samples.
Finally, we selected the top $K$ samples, which are determined to be the hardest by our initial model (i.e., the samples with probabilities of bogus source and real detection both tending toward 0.5), and annotate them, integrating the artificial experience. Although the initial model may not achieve high performance with only a few labeled samples, it guides us in identifying challenging samples, thereby reducing the cost of expert labeling. The annotation process is illustrated by the first arrow in Figure \ref{fig:sample}. 

Active learning stage: After obtaining the labels for the $K$ hard samples, we used the $(M + K)$ labeled data to train an active training model, which has the same structure as the initial model, during the ALS. Following the training phase, we infer the results for the remaining $(N-M-K)$ samples as well.  
Unlike the ITS, we extracted the higher dimension of each two-dimensional (2D) result as a confidence score. However, for the $(N_{\neg test} -M-K)$ 1-dimensional scores, we arrange them in descending order. 
As illustrated in the second arrow in Figure \ref{fig:sample}, We chose the top $V$ samples, which correspond to the highest confidence samples according to the active training model (i.e., the samples with probabilities of bogus detection or real source trending toward 1) to assign pseudo-labels. These pseudo-labels are predicted labels generated by the active training model.
The overall operation in the ALS is illustrated in the middle of Figure \ref{fig:pipeline}.
Although the active training model may not achieve satisfactory results with only $(M+K)$ labeled samples, it exhibits a certain discriminative ability. Therefore, the pseudo-labels of the samples with confidence scores surpassing a predetermined threshold $\tau$ can be considered correct judgments.

Semi-supervised learning stage: During the SSLS, the $(M+K+V)$ samples are employed to train a semi-supervised training model, as illustrated in the right portion of Figure \ref{fig:pipeline}. This model has been designed with the same structure as the initial model and the active training model. Following the training process, the $(N-M-K)$ unlabeled samples are sent to the semi-supervised training model to obtain the final predicted results. 
This stage can be iterated for $R$ times, as depicted in the right section of Figure \ref{fig:sample}. Specifically, after training the semi-supervised training model in the $r$-th iteration, we selected the samples from the $(N_{\neg test}-M-K)$ unlabeled data that surpass a predetermined threshold $\tau$ and assigned them pseudo-labels. Subsequently, we retrained the semi-supervised training model from scratch in the $(r+1)$-th iteration. By selecting the best $R$ iterations, we have been able to achieve competitive results. Algorithm \ref{alg:algorithm1} summarizes our RB-C1000. 

By adopting this comprehensive approach, we can leverage the strengths of both initial model training, active learning, and semi-supervised learning. This iterative process allows us to refine the model's performance and enhance the accuracy and reliability of the final classification results.

\section{Experiments} \label{sec:experiments}

In this section, we provide a brief overview of the experimental settings used in our experiments. We then describe the metric used for evaluating the performance of our classification task. Finally, we show the results of our real-bogus classifier, which leverages both active learning and semi-supervised learning techniques.

\subsection{Experimental settings} \label{subsec:setting}
We utilized three new complied datasets we collected to demonstrate the effectiveness of our method. Each dataset comprises 43000 samples, consisting of 13000 real sources and 30000 bogus detections.  Specifically, the ZTF-NEWg dataset, which consists of 43000 g-band triplet image stamps, and the ZTF-NEWr dataset, which comprises 43000 r-band triplet image stamps. Additionally, the ZTF-NEWm dataset contains a combination of 21500 g-band triplet image stamps and 21500 r-band triplet image stamps which are randomly sampled by the ZTF-NEWg and the ZTF-NEWr, respectively. Each triplet image stamp within the datasets consists of three components: a 63 pixels $\times$ 63 pixels science image stamp, the corresponding 63 pixels $\times$ 63 pixels reference image stamp, and the corresponding 63 pixels $\times$ 63 pixels difference image stamp. Here, each pixel corresponds to 1.012 arcseconds. 

In our active and semi-supervised setting, we allocated 10\% of each dataset for testing ($N_{test} = 4300$). The amount of initial labeled data and the count of challenging samples are $M=900$ and $K=100,$ respectively. We partitioned the $(M+K) = 1000$ samples into training and validation sets, with 60\% for training and 40\% for validation. The validation set serves as a tool for selecting model parameters during the training process of a real and bogus classification model. By evaluating the model's performance on the validation set, we can identify the model parameters that yield the best results on the validation set and save these optimal parameters as the final real and bogus classification model parameters.

We implemented our method based on the Pytorch \citep{pytorch} and adopted a Nvidia Tesla v100 GPU with a batch size of 256 for training. For the backbone architecture, we employed ResNet-18 \citep{He2015deep} and made appropriate modifications to adjust the output dimension for the classification of real and bogus transients. Each sample from the datasets has a size of $H \times W \times 3$, where $H$ and $W$ represent the height and width of image stamps respectively. Here, the number 3 represents the count of channels, corresponding to the science image stamp, the reference image stamp, and the difference image stamp, each contributing one channel.  We used $H=63$ pixels and $W=63$ pixels here. To adapt the images to the input dimension of ResNet-18, we interpolated all images to a uniform size of 224 pixels $\times$ 224 pixels $\times$ 3. To improve the flipping invariance, we applied two data augmentation operation schemes on the training samples. One is random horizontal flipping, the other is random vertical flipping. Both operations have a probability of 0.5. We train the model by cross entropy loss in 100 epochs with an Adam optimizer whose learning rate is 2e-4. The predetermined threshold $\tau$ is 95\%. We repeated the SSLS stage $R=3$ times. 

\begin{figure}[t]
\centering
\includegraphics[width=0.4\linewidth]{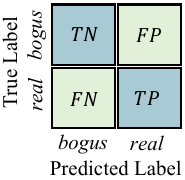} 
\caption{Definitions of $TN$, $FP$, $FN$, and $TP$.
\label{fig:con_m}}
\end{figure}

We have observed that achieving the classification of real and bogus transients is possible by solely utilizing the difference image stamp or by employing the two-channel image stamp  (the one-channel science image stamp and the one-channel reference image stamp). The existing approach \citep{Acero-Cuellar2022There} on the real-bogus classifier provides evidence that a higher number of channels leads to improved classification performance. In our approach, we aim to obtain more precise results during the early stage of time-domain surveys. Therefore, we conduct experiments in our method using three-channel images. A summary of the basic parameters is presented in Table \ref{ex_p}.

\subsection{Performance indicators} \label{subsec:Indicators}

The essence of the classification of real and bogus transients is a classification problem. Thus, we can evaluate the performance with precision, accuracy, recall, F1 score, Matthews correlation coefficient (MCC), and average precision (AP), which are commonly used in classification problems. 
AP is the area under the precision-recall (PR) curve, and the specific formulas for precision, accuracy, recall, F1 score, and MCC are as follows:

\begin{gather}
    {\rm Precision}=\frac{TP}{TP+FP},\\
    {\rm Accuracy}=\frac{TP+TN}{TP+FP+TN+FN},\\
    {\rm Recall}=\frac{TP}{TP+FN},\\
    \rm F1  \, score=\frac{2 \times Precision \times Recall}{Precision+Recall},\\
    {\rm MCC}=\frac{TP \times TN-FP \times FN}{\sqrt{(TP+FP)\times(TP+FN)\times(TN+FP)\times(TN+FN)}}.
\end{gather}
where $TP$ (true positive) indicates the real sources which are correctly classified as real and $FN$ (false negative) denotes the real sources that are classified as bogus; $FP$ (false positive) means the bogus detections that are classified as real and $TN$ (true negative) indicates the bogus detections which are correctly classified as bogus. The definitions of $TN$, $FP$, $FN$, and $TP$ are shown in Figure \ref{fig:con_m}.

Precision is the probability of real sources among the real sources judged by our network. Accuracy indicates the percentage of correctly classified in the total data. 
Recall represents the proportion of real sources in the whole dataset that are identified by our network.
F1 score and MCC are comprehensive evaluation indices. The closer all these performance indicators are to 1, the better the method's performance.

\begin{figure*}[t]
\centering
\includegraphics[width=0.95\linewidth]{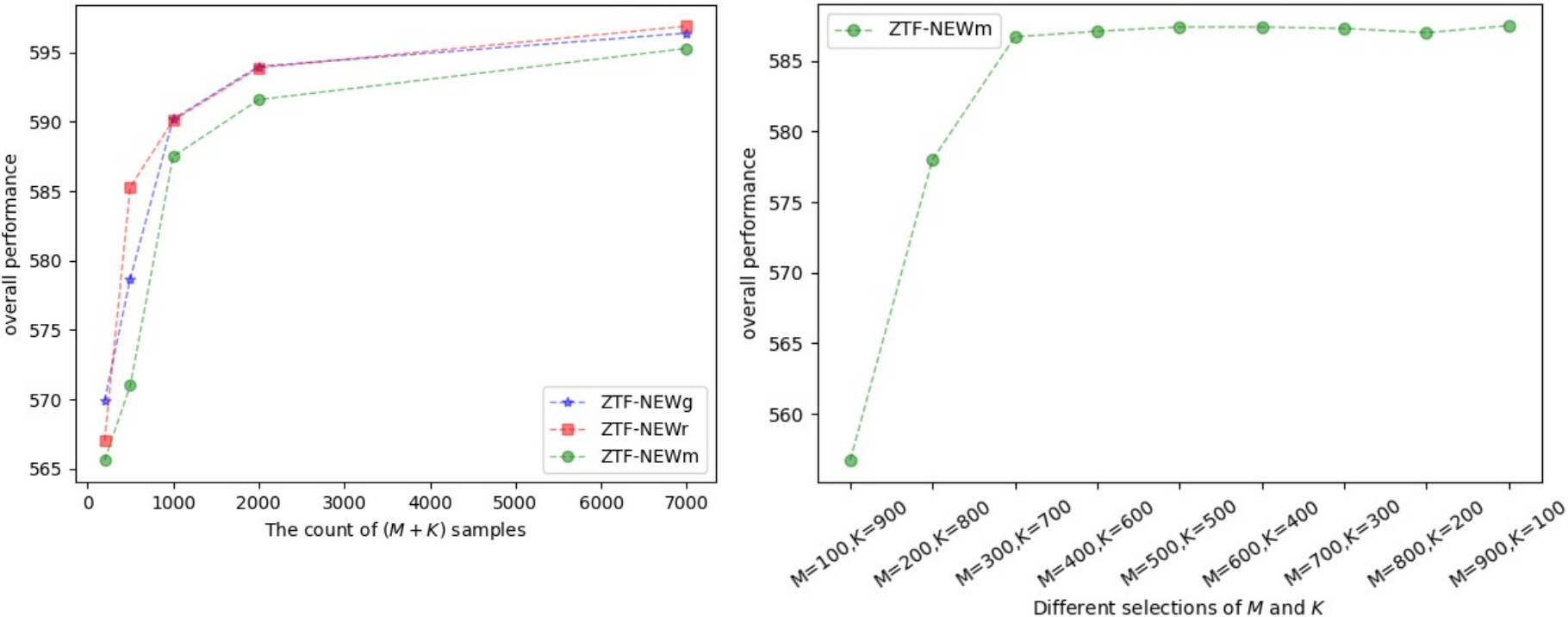} 
\caption{(a) Relationship between the quantity of labeled samples $(M + K)$ and the overall performance. (b) Relationship between different selections of $M$ and $K$ and the overall results in the ZTF-NEWm. Figures (a) and (b) only display the mean value of the overall performance.
\label{fig:M+K}}
\end{figure*}

\begin{figure*}[t]
\centering
\includegraphics[width=0.8\linewidth]{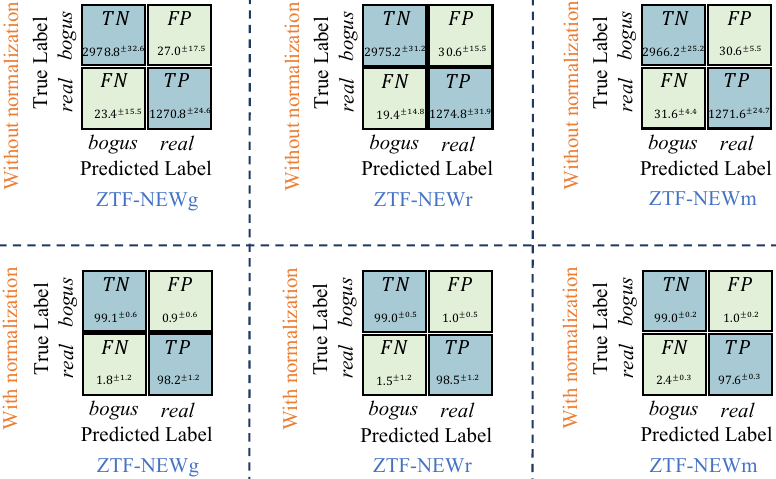} 
\caption{Results of confusion matrices using RB-C1000 on our three datasets. The top half of this figure represents the unnormalized results, which provide a direct representation of the raw classification outcomes. The lower half of this figure displays the normalized results.
\label{fig:matrix}}
\end{figure*}

\subsection{Test results} \label{subsec:Results}
We evaluated the performance of our method on three newly compiled datasets. To verify the effectiveness of our algorithm, each experiment was repeated five times. Both the mean and standard deviation of the results obtained over the five trials were then reported.

The performance results of our model are presented in Table \ref{tab:main_results}, where we offer the precision, accuracy, recall, F1 score, Matthews correlation coefficient (MCC), and average precision (AP) as performance measurement indicators. The overall performance measurement was obtained by summing the values of these six indicators, as depicted in the last column of Table \ref{tab:main_results}.
Taking the ZTF-NEWg dataset as an example, we see that in the initial training stage (ITS), using 900 labeled samples, our model achieved an overall performance of 572.2\%; the accuracy and recall were 96.7\% and 95.5\%, respectively. This result demonstrates that our model can effectively distinguish real detections and bogus sources with a limited number of labeled data. In the active learning stage (ALS), we incorporated an additional set of the $K$ challenging samples, which were determined by the model trained in the ITS. These $K$ samples were labeled by experts. By utilizing the $(M+K=1000)$ labeled samples for training in the ALS, the overall performance increased by 13.9\% compared to the overall result of the ITS. We note that the observed improvement of 13.9\% cannot be solely attributed to active learning, as increasing the amount of training data itself may also contribute to enhanced performance. The effect of active learning is depicted in the next paragraph  and the experiment of active learning is shown in Table \ref{tab:active}. In the semi-supervised learning stage (SSLS), we employed $V$ high-confidence samples, in addition to the labeled $(M+K)$ samples, to train the semi-supervised training model. By repeating the SSLS process for $R=1$, 2, 3, 4, and 5 iterations, we achieved respective overall performance indicators of 588.2\%, 589.9\%, 590.2\%, 589.8\%, and 589.9\%. This iterative process in the SSLS allowed us to refine the model's performance and enhance the accuracy and reliability of the final classification results. We notice that SSLS $(R = 3)$ often shows the best performance on the overall indicator on these three datasets we collected, so we set $R = 3$ in this paper and the remaining experiments (in Tables \ref{tab:active}, \ref{tab:count_unlabel}, \ref{tab:M+K}, \ref{tab:selectMandK}, and \ref{tab:transfer}) are all based on the condition $R = 3$.

Effect of active learning: 
We evaluated the improvement made by active learning in Table \ref{tab:active}. For each dataset, the results (given in rows 1 and  2) were obtained by randomly selecting $K$ samples and selecting $K$ hard samples from $(N_{\neg test}-M)$ unlabeled data, respectively. The overall performance result in the last column shows an improvement of 1.4\%, 3.4\%, and 7.2\% for the ZTF-NEWg, the ZTF-NEWr, and the ZTF-NEWm, respectively, by active learning. Therefore, by selectively annotating the most informative samples, active learning aims to achieve higher model performance by leveraging fewer labeled samples, thus reducing annotation costs and improving the overall learning process efficiency.

Effect of semi-supervised learning: 
We aim to quantify the contribution of semi-supervised learning. The results (presented in rows 2 and 5 of each dataset in Table \ref{tab:main_results}) demonstrate the gains achieved through the SSLS. We observe improvements of 4.1\%, 8.7\%, and 7.3\% in the overall performance of the three datasets,  respectively. These results indicate that the SSLS effectively utilizes the unlabeled data to enhance the model's performance, surpassing the performance obtained using only the limited labeled data. The iterative process employed in the SSLS allows for a continuous refinement of the model's performance, ultimately improving accuracy and reliability in the final classification results.

Effect of the count of unlabeled data usage:
As we know, the number of samples of each dataset is $N=43000$, where 10\% ($N_{test}=4300$) is reserved for testing. The count of labeled samples for initial training and the number of challenging samples are $M=900$ and $K=100,$ respectively. All other 33770 samples except $N_{test}$, $M$, and $K$ are unlabeled data. We explored the effect of the amount of unlabeled data usage on performance. 
We utilized 5000 samples (the remaining 28770 unlabeled samples are not used), 10000 samples (the remaining 23770 unlabeled samples are not used), 15000 samples (the remaining 18770 unlabeled samples are not used), and 20000 samples (the remaining 13770 unlabeled samples were not used) of the unlabeled 33770 samples. The results (depicted in Table \ref{tab:count_unlabel}) demonstrate the overall performance achieved on the ZTF-NEWg dataset, namely, is 589.7\%, 590.2\%, 589.6\%, and 586.1\%, respectively. Similarly, for the ZTF-NEWr dataset, we obtained the overall performance score of 589.7\%, 590.1\%, 590.5\%, and 588.1\%. The overall performance results are 587.5\%, 587.5\%, 586.7\%, and 587.0\% on the ZTF-NEWm dataset. 
For the ZTF-NEWg, utilizing 10000 unlabeled samples offered the best performance. Using 15000 unlabeled data offered the best overall indicator on the ZTF-NEWr. For the ZTF-NEWm, 5000 unlabeled samples or 10000 unlabeled samples achieved competitive results. It can be seen from experiments that 10000 unlabeled sample usage size often give the best results.
Note that for each additional usage of 5000 unlabeled data, the training time experiences an approximate increase of half an hour. Given this consideration, it is feasible to select either 5000 or 10000 unlabeled data usage. All other experimental results (presented in Tables \ref{tab:main_results}, \ref{tab:active}, \ref{tab:M+K}, \ref{tab:selectMandK}, and \ref{tab:transfer}) were obtained using the case of 10000 unlabeled data usage. This choice strikes a reasonable balance between leveraging the benefits of unlabeled data and managing the associated training time.

Effect of the labeled sample quantity $(M + K)$:
We explore the impact of the labeled sample quantity $(M+K)$ on performance, as presented in Table \ref{tab:M+K} and depicted in the left part of Figure \ref{fig:M+K}. It can be seen that no matter which dataset, an increasing quantity of $(M+K)$ leads to a progressive improvement in overall performance. The overall performance experiences a significant enhancement when the number of annotations rises from 200 to 500. However, once the annotation count surpasses 1000, the rate of improvement in overall performance begins to slow down. Therefore, we opt for an overall number of tags of $(M+K)$ = 1000, taking into consideration both the labeling cost and performance factors. All experiments presented in Tables \ref{tab:main_results}, \ref{tab:active}, \ref{tab:count_unlabel}, \ref{tab:selectMandK}, and \ref{tab:transfer} are all based on the condition $(M+K)$ = 1000.

Effect of stability on different selections of $M$ and $K$: 
We tested our model's stability on the ZTF-NEWm dataset as an example, which is the most challenging one of our three datasets. We explore different selections of $M$ and $K$, which represent the initial labeled quantity and most challenging labeled samples' quantity, respectively, as shown in Table \ref{tab:M+K} and Figure \ref{fig:M+K}. It can be shown that our algorithm's results tend to stabilize once $M\geq400$. The highest overall performance is achieved at ratios of 0.9, resulting in performances of 587.5\%. Therefore, we choose a ratio of 0.9, which is employed in Tables \ref{tab:main_results}, \ref{tab:active},  \ref{tab:count_unlabel}, \ref{tab:M+K}, and \ref{tab:transfer}.

The efficiency of our approach:
We explored the extent to which annotations from fully supervised techniques can match the performance of RB-C1000. We also used ZTF-NEWm as an example. We trained a fully supervised model using ResNet18 while maintaining the test set size equivalent to RB-C1000. We conducted fully supervised experiments with randomly labeled sets comprising 1000, 1500, 2000, 2500, and 3000 annotations, while maintaining the ratio of real to bogus samples, with the results  shown in Table \ref{tab:efficiency}. Our methodology yields results surpassing those of the fully supervised model employing 2500 annotations, despite using only 1000 labels available. The efficiency of our RB-C1000 is significantly higher than the random annotation of the fully supervised model.

Effect of bands on performance:
We evaluated the impact of bands on performance. 
Tables \ref{tab:main_results} and \ref{tab:active} demonstrate that the impact of our method on the ZTF-NEWm dataset is more pronounced compared to the single-band datasets. The transition from a single-band dataset to a dual-band dataset introduces higher diversity and complexity. Our approach proves to be more effective for challenging data.
Additionally, we conducted an evaluation of the performance when training on the g-band and r-band datasets separately, and then testing on the opposite band. Specifically, for each model trained on the g-band dataset, we randomly selected 10\% of the data from the r-band dataset as the test samples, and vice versa. The results (presented in Table \ref{tab:transfer}) demonstrate that the performance is satisfactory when training on the g-band and testing on the r-band; conversely, the performance is notably poorer. These results show the importance of not directly applying a model trained on one band to another band. The discrepancy in performance between training and testing on different bands suggests that the models trained on one specific band are not robust enough to be aptly generalized  to the other band.

Discussion: Our RB-C1000 is improved by active learning and semi-supervised learning. For the ZTF-NEWg, the ZTF-NEWr, and the ZTF-NEWm, the improvement made by active learning is 1.4\%, 3.4\%, and 7.2\% respectively and the improvement made by semi-supervised learning is 4.1\%, 8.7\%, and 7.3\% respectively. The efficiency of our RB-C1000 is significantly higher than the random annotation of the fully supervised model. Our methodology yields results surpassing those of the fully supervised model employing 2500 annotations, despite using only 1000 labels available.  We note that the performance of the second row of ZTF-NEWr in Table \ref{tab:main_results} is slightly worse than that of the first row. We realize that adding difficult samples may degrade the performance, but after adding active learning and semi-supervised learning, the advantage of difficult samples on performance is revealed again.

We present the results of the confusion matrices obtained using RB-C1000 on the three collected datasets, namely ZTF-NEWg, ZTF-NEWr, and ZTF-NEWm, in Figure \ref{fig:matrix}. The upper half of Figure \ref{fig:matrix} showcases the unnormalized results, while the lower half displays the normalized results. The unnormalized results provide a direct representation of the raw classification outcomes, while the normalized results offer a normalized view, accounting for the distribution of $TP$ (true positive), $FN$ (false negative), $FP$ (false positive), and $FN$ (false negative) predictions. It is important to note that, as mentioned in Section \ref{sec:data}, there may be a potential bias between our datasets and the complete data stream. Therefore, the performance observed on these datasets may not directly generalize to real-world scenarios involving general transient detection. Our RB-C1000, which incorporates active learning and semi-supervised learning techniques, demonstrates significant potential in effectively classifying real and bogus sources. Active learning empowers deep learning algorithms to actively acquire new labeled data, leading to improved model performance, reduced labeling effort, and more cost-effective data annotation. Semi-supervised learning provides an effective approach for leveraging large amounts of unlabeled data, improving model performance, and reducing labeling effort and costs. The active-learning technique and the semi-supervised learning technique are valuable techniques in situations where the availability of labeled data is limited.

\section{Summary and conclusion} \label{sec:conclusion}
In this paper, we propose RB-C1000, a novel deep-learning method using active learning and semi-supervised learning, which obtains a highly promising real-bogus classifier capability with only 1000 labels. The algorithm facilitates efficient and effective classification of transients in the early period of a time-domain survey. 
A module based on our method will be incorporated into the data processing pipeline of the upcoming 2.5-meter Wide-Field Survey Telescope (WFST) six-year survey, aimed at discovering and monitoring various transient events. 
To verify the applicability of our approach, we constructed three new datasets (ZTF-NEWg, ZTF-NEWr, and ZTF-NEWm) from the Zwicky Transient Facility (ZTF). Our RB-C1000 model achieves an average accuracy of 98.8\%, 98.8\%, and 98.6\% on these datasets, respectively. It's crucial to note that there may be a potential bias between our datasets and the complete data stream. Therefore, these resulting performances may not represent the performance of our approach in real-world sky surveys involving general transient detection. Our newly complied datasets only work in terms of testing our approach, as well as the effectiveness of each component such as the active learning and semi-supervised learning parts.
In practice, by using RB-C1000, we can significantly reduce the consumption of human annotation and obtain a competitive model by leveraging training samples with only 1000 labels available. We adopted a Nvidia Tesla v100 GPU with a batch size of 256 for training. The total time taken for the training procedure was approximately 1.5 hours.
RB-C1000 is a flexible algorithm that can also be applied to next-generation telescopes, such as the Vera C. Rubin Observatory Legacy Survey of Space and Time (LSST), to address the real-source detection issue in the early stage of a time-domain survey. 

\section*{Data availability}
The code and three datasets we collected are available at \url{https://github.com/cherry0116/RB-C1000}. 

\begin{acknowledgements}
     We are grateful to the anonymous referee for providing many useful comments. This work is supported by the National Key Research and Development Program of China (2023YFA1608100) and the Strategic Priority Research Program of the Chinese Academy of Sciences, Grant No. XDB 41000000. We gratefully acknowledge the support of the National Natural Science Foundation of China (NSFC, grant No. 12173037, 12233008), the CAS Project for Young Scientists in Basic Research (No. YSBR-092),  the Fundamental Research Funds for the Central Universities (WK3440000006), and Cyrus Chun Ying Tang Foundations.
\end{acknowledgements}

%
%

\begin{appendix}

{\section{Additional tables}

\begin{table}[hb]
\onecolumn
\renewcommand\arraystretch{1.23}
\begin{center}
\caption{Performance on three datasets we collected.}\label{tab:main_results}
\begin{tabular}{c|p{4cm}<{\centering}|cccccc|p{2cm}<{\centering}} 
\hline
Row & Stage in our method & Precision & Accuracy & Recall & F1 score & MCC & AP & Sum \\
\hline
\hline
\multicolumn{9}{c}{ZTF-NEWg}\\
\hline
1 & ITS $(M=900)$ & ${93.7^{\pm{3.8}}}$ & ${96.7^{\pm{1.5}}}$ &$ {95.5^{\pm{2.5}}}$ &${94.6^{\pm{2.4}}}$ & ${92.2^{\pm{3.5}}}$ & ${99.6^{\pm{0.2}}}$ & ${572.2^{\pm{6.4}}}$
\\
2 & ALS $(M+K=1000)$ & ${97.3^{\pm{1.1}}}$ & ${98.4^{\pm{0.7}}}$ &$ {97.2^{\pm{1.4}}}$ &${97.3^{\pm{1.1}}}$ & ${96.1^{\pm{1.6}}}$ & ${ 99.9^{\pm{0.1}}}$ & ${586.1^{\pm{2.7}}}$
\\
3 & SSLS $(R=1)$ & ${97.8^{\pm{1.8}}}$ & ${98.6^{\pm{0.8}}}$ &$ {97.6^{\pm{1.1}}}$ &${97.7^{\pm{1.4}}}$ & ${96.7^{\pm{2.0}}}$ & ${99.8^{\pm{0.2}}}$ & ${588.2^{\pm{3.3}}}$
\\
4 & SSLS $(R=2)$ & ${97.4^{\pm{1.2}}}$ & ${98.8^{\pm{0.5}}}$ &$ { 98.7^{\pm{0.7}}}$ &${98.0^{\pm{0.9}}}$ & ${97.2^{\pm{1.2}}}$ & ${99.8^{\pm{0.2}}}$ & ${589.9^{\pm{2.1}}}$
\\
5 & SSLS $(R=3)$ & ${98.0^{\pm{1.3}}}$ & ${98.8^{\pm{0.3}}}$ &$ {98.2^{\pm{1.2}}}$ &${98.1^{\pm{0.5}}}$ & ${97.2^{\pm{0.8}}}$ & ${99.9^{\pm{0.1}}}$ & ${590.2^{\pm{2.0}}}$
\\
6 & SSLS $(R=4)$ & ${97.3^{\pm{1.7}}}$ & ${98.8^{\pm{0.5}}}$ &$ {98.7^{\pm{0.6}}}$ &${98.0^{\pm{0.8}}}$ & ${97.1^{\pm{1.2}}}$ & ${99.9^{\pm{0.1}}}$ & ${589.8^{\pm{2.3}}}$
\\
7 & SSLS $(R=5)$ & ${97.6^{\pm{1.9}}}$ & ${98.8^{\pm{0.5}}}$ &$ {98.5^{\pm{0.7}}}$ &${98.0^{\pm{0.7}}}$ & ${97.2^{\pm{1.0}}}$ & ${99.8^{\pm{0.2}}}$ & ${589.9^{\pm{2.4}}}$
\\
\hline
\hline
\multicolumn{9}{c}{ZTF-NEWr}\\
\hline
1 & ITS $(M=900)$ & ${95.6^{\pm{1.4}}}$ & ${97.8^{\pm{0.3}}}$ &$ {97.3^{\pm{1.0}}}$ &${96.4^{\pm{0.5}}}$ & ${94.9^{\pm{0.7}}}$ & ${99.6^{\pm{0.3}}}$ & ${581.7^{\pm{2.0}}}$
\\
2 & ALS $(M+K=1000)$ & ${96.0^{\pm{1.0}}}$ & ${97.8^{\pm{0.5}}}$ &$ {96.8^{\pm{2.7}}}$ &${96.4^{\pm{0.9}}}$ & ${94.8^{\pm{1.3}}}$ & ${99.6^{\pm{0.3}}}$ & ${581.4^{\pm{3.3}}}$
\\
3 & SSLS $(R=1)$ & ${96.9^{\pm{1.2}}}$ & ${98.4^{\pm{0.3}}}$ &$ {97.7^{\pm{1.5}}}$ &${97.3^{\pm{0.5}}}$ & ${96.1^{\pm{0.7}}}$ & ${99.6^{\pm{0.4}}}$ & ${586.0^{\pm{2.2}}}$
\\
4 & SSLS $(R=2)$ & ${97.0^{\pm{1.1}}}$ & ${98.5^{\pm{0.5}}}$ &$ {98.1^{\pm{1.2}}}$ &${97.5^{\pm{0.8}}}$ & ${96.5^{\pm{1.1}}}$ & ${99.7^{\pm{0.4}}}$ & ${587.4^{\pm{2.2}}}$
\\
5 & SSLS $(R=3)$ & ${97.7^{\pm{1.1}}}$ & ${98.8^{\pm{0.2}}}$ &$ {98.5^{\pm{1.2}}}$ &${98.1^{\pm{0.3}}}$ & ${97.3^{\pm{0.5}}}$ & ${99.7^{\pm{0.4}}}$ & ${590.1^{\pm{1.8}}}$
\\
6 & SSLS $(R=4)$ & ${97.6^{\pm{0.5}}}$ & ${98.9^{\pm{0.1}}}$ &$ {98.7^{\pm{0.5}}}$ &${98.1^{\pm{0.2}}}$ & ${97.3^{\pm{0.3}}}$ & ${99.7^{\pm{0.4}}}$ & ${590.2^{\pm{0.9}}}$
\\
7 & SSLS $(R=5)$ & ${97.4^{\pm{1.0}}}$ & ${98.7^{\pm{0.2}}}$ &$ {98.3^{\pm{0.6}}}$ &${97.9^{\pm{0.3}}}$ & ${97.0^{\pm{0.4}}}$ & ${99.6^{\pm{0.4}}}$ & ${588.9^{\pm{1.4}}}$
\\
\hline
\hline
\multicolumn{9}{c}{ZTF-NEWm}\\
\hline
1 & ITS $(M=900)$ & ${91.0^{\pm{2.8}}}$ & ${96.0^{\pm{0.8}}}$ &$ {96.5^{\pm{0.8}}}$ &${93.6^{\pm{1.2}}}$ & ${90.8^{\pm{1.7}}}$ & ${99.2^{\pm{0.3}}}$ & ${567.2^{\pm{3.7}}}$
\\
2 & ALS $(M+K=1000)$ & ${95.7^{\pm{1.6}}}$ & ${97.7^{\pm{0.6}}}$ &$ {96.6^{\pm{1.5}}}$ &${96.2^{\pm{1.0}}}$ & ${94.5^{\pm{1.5}}}$ & ${99.5^{\pm{0.4}}}$ & ${580.2^{\pm{2.9}}}$
\\
3 & SSLS $(R=1)$ & ${96.7^{\pm{0.7}}}$ & ${98.3^{\pm{0.2}}}$ &$ {97.8^{\pm{0.4}}}$ &${97.2^{\pm{0.3}}}$ & ${96.0^{\pm{0.4}}}$ & ${99.5^{\pm{0.5}}}$ & ${585.6^{\pm{1.1}}}$
\\
4 & SSLS $(R=2)$ & ${97.3^{\pm{1.0}}}$ & ${98.5^{\pm{0.3}}}$ &$ {97.7^{\pm{0.3}}}$ &${97.5^{\pm{0.4}}}$ & ${96.4^{\pm{0.6}}}$ & ${99.4^{\pm{0.7}}}$ & ${586.7^{\pm{1.5}}}$
\\
5 & SSLS $(R=3)$ & ${97.6^{\pm{0.4}}}$ & ${98.6^{\pm{0.1}}}$ &$ {97.6^{\pm{0.3}}}$ &${97.6^{\pm{0.2}}}$ & ${96.6^{\pm{0.3}}}$ & ${99.5^{\pm{0.6}}}$ & ${587.5^{\pm{0.8}}}$
\\
6 & SSLS $(R=4)$ & ${97.1^{\pm{1.5}}}$ & ${98.5^{\pm{0.4}}}$ &$ {98.1^{\pm{0.7}}}$ &${97.6^{\pm{0.6}}}$ & ${96.5^{\pm{0.8}}}$ & ${99.4^{\pm{0.5}}}$ & ${587.1^{\pm{2.0}}}$
\\
7 & SSLS $(R=5)$ & ${97.6^{\pm{0.7}}}$ & ${98.6^{\pm{0.2}}}$ &$ {97.7^{\pm{0.7}}}$ &${97.7^{\pm{0.4}}}$ & ${96.6^{\pm{0.5}}}$ & ${99.3^{\pm{0.8}}}$ & ${587.5^{\pm{1.4}}}$
\\
\hline
\hline
\end{tabular}
\tablefoot{We employ precision, accuracy, recall, F1 score, Matthews correlation coefficient (MCC), and average precision (AP) as performance measurement indicators. The overall performance measurement is obtained by summing the values of these six indicators (shown in the last column).}
\end{center}
\end{table}

\begin{table}[hb]
\renewcommand\arraystretch{1.2}
\begin{center}
\caption{Improved results through active learning.}\label{tab:active}
\begin{tabular}{p{4cm}<{\centering}|cccccc|p{2cm}<{\centering}} 
\hline
Select the $K$ hard samples & Precision & Accuracy & Recall & F1 score & MCC & AP & Sum \\
\hline
\hline
\multicolumn{8}{c}{ZTF-NEWg}\\
\hline
  & ${97.8^{\pm{1.1}}}$ & ${98.7^{\pm{0.5}}}$ &$ {97.8^{\pm{1.1}}}$ &${97.8^{\pm{0.9}}}$ & ${96.9^{\pm{1.2}}}$ & ${99.9^{\pm{0.1}}}$ & ${588.8^{\pm{2.2}}}$
\\
\checkmark & ${98.0^{\pm{1.3}}}$ & ${98.8^{\pm{0.3}}}$ &$ {98.2^{\pm{1.2}}}$ &${98.1^{\pm{0.5}}}$ & ${97.2^{\pm{0.8}}}$ & ${99.9^{\pm{0.1}}}$ & ${590.2^{\pm{2.0}}}$
\\
\hline
\hline
\multicolumn{8}{c}{ZTF-NEWr}\\
\hline
 & ${97.4^{\pm{1.8}}}$ & ${98.5^{\pm{0.4}}}$ &$ {97.5^{\pm{1.6}}}$ &${97.4^{\pm{0.6}}}$ & ${96.3^{\pm{0.8}}}$ & ${99.6^{\pm{0.4}}}$ & ${586.7^{\pm{2.7}}}$
\\
\checkmark & ${97.7^{\pm{1.1}}}$ & ${98.8^{\pm{0.2}}}$ &$ {98.5^{\pm{1.2}}}$ &${98.1^{\pm{0.3}}}$ & ${97.3^{\pm{0.5}}}$ & ${99.7^{\pm{0.4}}}$ & ${590.1^{\pm{1.8}}}$
\\
\hline
\hline
\multicolumn{8}{c}{ZTF-NEWm}\\
\hline
 & ${95.4^{\pm{2.5}}}$ & ${97.7^{\pm{0.7}}}$ &$ {97.0^{\pm{0.8}}}$ &${96.2^{\pm{1.1}}}$ & ${94.5^{\pm{1.6}}}$ & ${99.6^{\pm{0.2}}}$ & ${580.3^{\pm{3.3}}}$
\\
\checkmark & ${97.6^{\pm{0.4}}}$ & ${98.6^{\pm{0.1}}}$ &$ {97.6^{\pm{0.3}}}$ &${97.6^{\pm{0.2}}}$ & ${96.6^{\pm{0.3}}}$ & ${99.5^{\pm{0.6}}}$ & ${587.5^{\pm{0.8}}}$
\\
\hline
\hline
\end{tabular}
\tablefoot{We employ precision, accuracy, recall, F1 score, Matthews correlation coefficient (MCC), and average precision (AP) as performance measurement indicators. The overall performance measurement is obtained by summing the values of these six indicators (shown in the last column). All results listed are SSLS $(R = 3)$.}
\end{center}
\end{table}

\begin{table}[hb]
\renewcommand\arraystretch{1.4}
\centering
\begin{center}
\caption{Effect of the amount of unlabeled data usage.}\label{tab:count_unlabel}
\begin{tabular}{c|cccccc|p{1.5cm}<{\centering}|p{1.5cm}<{\centering}} 
\hline
The count of unlabeled data usage & Precision & Accuracy & Recall & F1 score & MCC & AP & Sum & Time\\
\hline
\hline
\multicolumn{9}{c}{ZTF-NEWg}\\
\hline
5000 & ${97.8^{\pm{1.1}}}$ & ${98.8^{\pm{0.4}}}$ &$ {98.1^{\pm{1.4}}}$ &${98.0^{\pm{0.6}}}$ & ${97.1^{\pm{0.9}}}$ & ${99.9^{\pm{0.1}}}$ & ${589.7^{\pm{2.1}}}$ & ${57.0^{\pm{1.6}}}$
\\
10000 & ${98.0^{\pm{1.3}}}$ & ${98.8^{\pm{0.3}}}$ &$ {98.2^{\pm{1.2}}}$ &${98.1^{\pm{0.5}}}$ & ${97.2^{\pm{0.8}}}$ & ${99.9^{\pm{0.1}}}$ & ${{\bf590.2}^{\pm{2.0}}}$ & ${81.0^{\pm{7.1}}}$
\\
15000 & ${97.4^{\pm{1.7}}}$ & ${98.8^{\pm{0.4}}}$ &${98.5^{\pm{0.8}}}$ &${98.0^{\pm{0.6}}}$ & ${97.1^{\pm{0.9}}}$ & ${99.9^{\pm{0.1}}}$ & ${589.6^{\pm{2.2}}}$  & ${120.6^{\pm{1.4}}}$
\\
20000 & ${96.1^{\pm{2.9}}}$ & ${98.3^{\pm{0.9}}}$ &${98.5^{\pm{0.7}}}$ &${97.3^{\pm{1.4}}}$ & ${96.1^{\pm{2.0}}}$ & ${99.7^{\pm{0.2}}}$ & ${586.1^{\pm{4.0}}}$ & ${166.2^{\pm{15.8}}}$
\\
\hline
\hline
\multicolumn{9}{c}{ZTF-NEWr}\\
\hline
5000 & ${97.8^{\pm{1.1}}}$ & ${98.8^{\pm{0.3}}}$ &$ {98.2^{\pm{0.5}}}$ &${98.0^{\pm{0.5}}}$ & ${97.1^{\pm{0.6}}}$ & ${99.7^{\pm{0.3}}}$ & ${589.7^{\pm{1.5}}}$ & ${57.0^{\pm{2.0}}}$
\\
10000 & ${97.7^{\pm{1.1}}}$ & ${98.8^{\pm{0.2}}}$ &$ {98.5^{\pm{1.2}}}$ &${98.1^{\pm{0.3}}}$ & ${97.3^{\pm{0.5}}}$ & ${99.7^{\pm{0.4}}}$ & ${590.1^{\pm{1.8}}}$ & ${86.2^{\pm{11.3}}}$
\\
15000 & ${97.6^{\pm{1.5}}}$ & ${98.9^{\pm{0.4}}}$ &${98.8^{\pm{1.1}}}$ &${98.2^{\pm{0.7}}}$ & ${97.4^{\pm{1.0}}}$ & ${99.7^{\pm{0.5}}}$ & ${590.5^{\pm{2.3}}}$ & ${119.4^{\pm{4.2}}}$
\\
20000 & ${97.4^{\pm{0.9}}}$ & ${98.6^{\pm{0.3}}}$ &$ {98.0^{\pm{1.0}}}$ &${97.7^{\pm{0.6}}}$ & ${96.7^{\pm{0.8}}}$ & ${99.6^{\pm{0.6}}}$ & ${588.1^{\pm{1.8}}}$ & ${157.2^{\pm{4.1}}}$
\\
\hline
\hline
\multicolumn{9}{c}{ZTF-NEWm}\\
\hline
5000 & ${97.0^{\pm{0.4}}}$ & ${98.5^{\pm{0.2}}}$ &$ {98.1^{\pm{0.5}}}$ &${97.6^{\pm{0.2}}}$ & ${96.5^{\pm{0.4}}}$ & ${99.7^{\pm{0.2}}}$ & ${587.5^{\pm{0.8}}}$ & ${60.2^{\pm{1.6}}}$
\\
10000 & ${97.6^{\pm{0.4}}}$ & ${98.6^{\pm{0.1}}}$ &$ {97.6^{\pm{0.3}}}$ &${97.6^{\pm{0.2}}}$ & ${96.6^{\pm{0.3}}}$ & ${99.5^{\pm{0.6}}}$ & ${587.5^{\pm{0.8}}}$ & ${90.0^{\pm{2.8}}}$
\\
15000 & ${96.5^{\pm{1.4}}}$ & ${98.4^{\pm{0.4}}}$ &${98.4^{\pm{0.5}}}$ &${97.4^{\pm{0.6}}}$ & ${96.3^{\pm{0.9}}}$ & ${99.6^{\pm{0.5}}}$ & ${586.7^{\pm{2.0}}}$ & ${114.6^{\pm{1.2}}}$
\\
20000 & ${97.5^{\pm{0.6}}}$ & ${98.5^{\pm{0.2}}}$ &$ {97.5^{\pm{0.9}}}$ &${97.5^{\pm{0.3}}}$ & ${96.4^{\pm{0.4}}}$ & ${99.7^{\pm{0.2}}}$ & ${587.0^{\pm{1.2}}}$ & ${151.0^{\pm{1.4}}}$
\\
\hline
\hline
\end{tabular}
\tablefoot{We employed precision, accuracy, recall, F1 score, Matthews correlation coefficient (MCC), and average precision (AP) as performance measurement indicators. The overall performance measurement is obtained by summing the values of these six indicators (shown in the second last column). The consumption time for each setting (The count of unlabeled samples usage are 5000, 10000, 15000, and 20000) is provided in the last column where the unit used is minute. All results listed are SSLS $(R = 3)$.}
\end{center}
\end{table}

\begin{table}[hb]
\renewcommand\arraystretch{1.4}
\centering
\begin{center}
\caption{Effect of our model's stability by different selections of $M$ and $K$.}
\label{tab:selectMandK}
\begin{tabular}{c|cccccc|p{2cm}<{\centering}} 
\hline
Different selections of $M$ and $K$ & Precision & Accuracy & Recall & F1 score & MCC & AP & Sum\\
\hline
\hline
\multicolumn{8}{c}{ZTF-NEWm}\\
\hline
$M$=100, $K$=900 & ${96.2^{\pm{6.0}}}$ & ${94.9^{\pm{2.8}}}$ &$ {87.1^{\pm{11.9}}}$ &${90.8^{\pm{5.5}}}$ & ${88.0^{\pm{6.5}}}$ & ${99.8^{\pm{0.3}}}$ & ${556.7^{\pm{16.1}}}$
\\
$M$=200, $K$=800 & ${96.2^{\pm{3.4}}}$ & ${97.4^{\pm{1.4}}}$ &$ {95.2^{\pm{5.5}}}$ &${95.6^{\pm{2.5}}}$ & ${93.9^{\pm{3.3}}}$ & ${99.7^{\pm{0.3}}}$ & ${578.0^{\pm{7.8}}}$
\\
$M$=300, $K$=700 & ${98.3^{\pm{1.0}}}$ & ${98.4^{\pm{0.4}}}$ &$ {96.5^{\pm{1.7}}}$ &${97.4^{\pm{0.7}}}$ & ${96.3^{\pm{0.9}}}$ & ${99.9^{\pm{0.1}}}$ & ${586.7^{\pm{2.3}}}$
\\
$M$=400, $K$=600 &${97.4^{\pm{0.7}}}$ & ${98.5^{\pm{0.2}}}$ &$ {97.6^{\pm{1.1}}}$ &${97.5^{\pm{0.2}}}$ & ${96.4^{\pm{0.4}}}$ & ${99.7^{\pm{0.1}}}$ & ${587.1^{\pm{1.3}}}$
\\
$M$=500, $K$=500 & ${97.3^{\pm{0.9}}}$ & ${98.5^{\pm{0.3}}}$ &$ {97.8^{\pm{0.5}}}$ &${97.5^{\pm{0.5}}}$ & ${96.5^{\pm{0.8}}}$ & ${99.9^{\pm{0.1}}}$ & ${587.4^{\pm{1.5}}}$
\\
$M$=600, $K$=400 & ${97.6^{\pm{0.8}}}$ & ${98.5^{\pm{0.5}}}$ &$ {97.5^{\pm{1.7}}}$ &${97.5^{\pm{0.8}}}$ & ${96.5^{\pm{1.2}}}$ & ${99.7^{\pm{0.2}}}$ & ${587.4^{\pm{2.4}}}$
\\
$M$=700, $K$=300 & ${97.7^{\pm{0.9}}}$ & ${98.5^{\pm{0.3}}}$ &$ {97.5^{\pm{0.8}}}$ &${97.6^{\pm{0.5}}}$ & ${96.5^{\pm{0.7}}}$ & ${99.5^{\pm{0.4}}}$ & ${587.3^{\pm{1.6}}}$
\\
$M$=800, $K$=200 & ${97.3^{\pm{1.1}}}$ & ${98.5^{\pm{0.2}}}$ &$ {97.7^{\pm{1.4}}}$ &${97.5^{\pm{0.4}}}$ & ${96.4^{\pm{0.6}}}$ & ${99.6^{\pm{0.4}}}$ & ${587.0^{\pm{2.0}}}$
\\
$M$=900, $K$=100 & ${97.6^{\pm{0.4}}}$ & ${98.6^{\pm{0.1}}}$ &$ {97.6^{\pm{0.3}}}$ &${97.6^{\pm{0.2}}}$ & ${96.6^{\pm{0.3}}}$ & ${99.5^{\pm{0.6}}}$ & ${587.5^{\pm{0.8}}}$
\\
\hline
\end{tabular}
\tablefoot{It can be shown that our algorithm's results tend to stabilize once $M\geq400$. We employ precision, accuracy, recall, F1 score, Matthews correlation coefficient (MCC), and average precision (AP) as performance measurement indicators. The overall performance measurement is obtained by summing the values of these six indicators (shown in the last column). All results listed are SSLS $(R = 3)$.}
\end{center}
\end{table}

\begin{table}[hb]
\centering
\renewcommand\arraystretch{1.3}
\caption{Impact of the labeled sample quantity $(M + K)$.}
\label{tab:M+K}
\begin{tabular}{c|cccccc|p{2cm}<{\centering}} 
\hline
The count of $(M+K)$ samples & Precision & Accuracy & Recall & F1 score & MCC & AP & Sum\\
\hline
\hline
\multicolumn{8}{c}{ZTF-NEWg}\\
\hline
200 & ${92.5^{\pm{4.4}}}$ & ${96.3^{\pm{0.8}}}$ &$ {95.9^{\pm{2.9}}}$ &${94.1^{\pm{1.2}}}$ & ${91.6^{\pm{1.7}}}$ & ${99.5^{\pm{0.2}}}$ & ${569.9^{\pm{5.7}}}$
\\
500 & ${96.2^{\pm{3.6}}}$ & ${97.5^{\pm{1.2}}}$ &$ {95.5^{\pm{3.7}}}$ &${95.8^{\pm{2.0}}}$ & ${94.1^{\pm{2.8}}}$ & ${99.7^{\pm{0.3}}}$ & ${578.7^{\pm{6.3}}}$
\\
1000 & ${98.0^{\pm{1.3}}}$ & ${98.8^{\pm{0.3}}}$ &$ {98.2^{\pm{1.2}}}$ &${98.1^{\pm{0.5}}}$ & ${97.2^{\pm{0.8}}}$ & ${99.9^{\pm{0.1}}}$ & ${590.2^{\pm{2.0}}}$
\\
2000 & ${98.9^{\pm{0.6}}}$ & ${99.3^{\pm{0.2}}}$ &$ {98.8^{\pm{1.0}}}$ &${98.8^{\pm{0.3}}}$ & ${98.3^{\pm{0.4}}}$ & ${99.8^{\pm{0.3}}}$ & ${594.0^{\pm{1.3}}}$
\\
7000 & ${99.3^{\pm{0.4}}}$ & ${99.6^{\pm{0.1}}}$ &${99.3^{\pm{0.4}}}$ &${99.3^{\pm{0.2}}}$ & ${99.0^{\pm{0.2}}}$ & ${100.0^{\pm{0.0}}}$ & ${596.4^{\pm{0.6}}}$
\\
\hline
\hline
\multicolumn{8}{c}{ZTF-NEWr}\\
\hline
200 & ${95.7^{\pm{3.4}}}$ & ${96.1^{\pm{0.8}}}$ &$ {91.5^{\pm{4.6}}}$ &${93.4^{\pm{1.4}}}$ & ${90.9^{\pm{1.9}}}$ & ${99.4^{\pm{0.4}}}$ & ${567.0^{\pm{6.2}}}$
\\
500 & ${97.2^{\pm{1.4}}}$ & ${98.3^{\pm{0.3}}}$ &$ {97.1^{\pm{1.0}}}$ &${97.1^{\pm{0.5}}}$ & ${95.9^{\pm{0.7}}}$ & ${99.7^{\pm{0.2}}}$ & ${585.3^{\pm{1.9}}}$
\\
1000 & ${97.7^{\pm{1.1}}}$ & ${98.8^{\pm{0.2}}}$ &$ {98.5^{\pm{1.2}}}$ &${98.1^{\pm{0.3}}}$ & ${97.3^{\pm{0.5}}}$ & ${99.7^{\pm{0.4}}}$ & ${590.1^{\pm{1.8}}}$
\\
2000 & ${98.6^{\pm{0.6}}}$ & ${99.3^{\pm{0.1}}}$ &$ {99.0^{\pm{0.5}}}$ &${98.8^{\pm{0.2}}}$ & ${98.3^{\pm{0.3}}}$ & ${99.8^{\pm{0.3}}}$ & ${593.9^{\pm{1.0}}}$
\\
7000 & ${99.4^{\pm{0.2}}}$ & ${99.6^{\pm{0.0}}}$ &${99.4^{\pm{0.2}}}$ &${99.4^{\pm{0.1}}}$ & ${99.1^{\pm{0.1}}}$ & ${100.0^{\pm{0.0}}}$ & ${596.9^{\pm{0.3}}}$
\\
\hline
\hline
\multicolumn{8}{c}{ZTF-NEWm}\\
\hline
200 & ${92.4^{\pm{2.2}}}$ & ${95.9^{\pm{0.4}}}$ &$ {94.3^{\pm{1.5}}}$ &${93.3^{\pm{0.5}}}$ & ${90.4^{\pm{0.8}}}$ & ${99.3^{\pm{0.2}}}$ & ${565.6^{\pm{2.9}}}$
\\
500 & ${91.6^{\pm{3.0}}}$ & ${96.5^{\pm{0.8}}}$ &$ {97.5^{\pm{0.9}}}$ &${94.4^{\pm{1.2}}}$ & ${92.0^{\pm{1.7}}}$ & ${99.0^{\pm{0.5}}}$ & ${571.0^{\pm{3.9}}}$
\\
1000 & ${97.6^{\pm{0.4}}}$ & ${98.6^{\pm{0.1}}}$ &$ {97.6^{\pm{0.3}}}$ &${97.6^{\pm{0.2}}}$ & ${96.6^{\pm{0.3}}}$ & ${99.5^{\pm{0.6}}}$ & ${587.5^{\pm{0.8}}}$
\\
2000 & ${98.2^{\pm{0.6}}}$ & ${99.0^{\pm{0.2}}}$ &$ {98.5^{\pm{0.4}}}$ &${98.4^{\pm{0.4}}}$ & ${97.6^{\pm{0.6}}}$ & ${99.9^{\pm{0.1}}}$ & ${591.6^{\pm{1.1}}}$
\\
7000 & ${99.2^{\pm{0.5}}}$ & ${99.4^{\pm{0.1}}}$ &${99.0^{\pm{0.4}}}$ &${99.1^{\pm{0.2}}}$ & ${98.7^{\pm{0.3}}}$ & ${100.0^{\pm{0.0}}}$ & ${595.3^{\pm{0.8}}}$
\\
\hline
\hline
\end{tabular}
\tablefoot{We employ precision, accuracy, recall, F1 score, Matthews correlation coefficient (MCC), and average precision (AP) as performance measurement indicators. The overall performance measurement is obtained by summing the values of these six indicators (shown in the last column). All results listed are SSLS $(R = 3)$.}
\end{table}

\begin{table}[hb]
\renewcommand\arraystretch{1.3}
\centering
\begin{center}
\caption{Efficiency of our approach.}
\label{tab:efficiency}
\begin{tabular}{c|cccccc|p{2cm}<{\centering}} 
\hline
annotation count & Precision & Accuracy & Recall & F1 score & MCC & AP & Sum\\
\hline
\hline
\multicolumn{8}{c}{ZTF-NEWm}\\
\hline
1000 & ${94.8^{\pm{1.5}}}$ & ${97.4^{\pm{0.6}}}$ &$ {96.6^{\pm{1.1}}}$ &${95.7^{\pm{0.9}}}$ & ${93.8^{\pm{1.3}}}$ & ${99.4^{\pm{0.3}}}$ & ${577.7^{\pm{2.5}}}$
\\
1500 & ${95.5^{\pm{1.8}}}$ & ${97.7^{\pm{0.6}}}$ &$ {97.1^{\pm{0.4}}}$ &${96.3^{\pm{1.0}}}$ & ${94.7^{\pm{1.5}}}$ & ${99.6^{\pm{0.2}}}$ & ${580.9^{\pm{2.6}}}$
\\
2000 & ${96.8^{\pm{0.9}}}$ & ${98.2^{\pm{0.5}}}$ &$ {97.1^{\pm{0.8}}}$ &${97.0^{\pm{0.8}}}$ & ${95.6^{\pm{1.1}}}$ & ${99.6^{\pm{0.2}}}$ & ${584.3^{\pm{1.9}}}$
\\
2500 & ${96.6^{\pm{1.3}}}$ & ${98.4^{\pm{0.5}}}$ &$ {98.1^{\pm{0.4}}}$ &${97.3^{\pm{0.8}}}$ & ${96.2^{\pm{1.1}}}$ & ${99.6^{\pm{0.4}}}$ & ${586.3^{\pm{2.0}}}$
\\
3000 & ${98.0^{\pm{0.7}}}$ & ${98.8^{\pm{0.2}}}$ &$ {97.9^{\pm{0.6}}}$ &${98.0^{\pm{0.3}}}$ & ${97.1^{\pm{0.5}}}$ & ${99.9^{\pm{0.1}}}$ & ${589.7^{\pm{1.1}}}$
\\
\hline
\end{tabular}
\tablefoot{Our methodology yields results surpassing those of the fully supervised model employing 2500 annotations, despite using only 1000 labels available. We employ precision, accuracy, recall, F1 score, Matthews correlation coefficient (MCC), and average precision (AP) as performance measurement indicators. The overall performance measurement is obtained by summing the values of these six indicators (shown in the last column).}
\end{center}
\end{table}

\begin{table}[hb]
\renewcommand\arraystretch{1.3}
\begin{center}
\caption{Evaluation of the performance when training on the g-band and r-band datasets separately, and then testing on the opposite band.}\label{tab:transfer}
\begin{tabular}{p{4cm}<{\centering}|cccccc|p{2cm}<{\centering}} 
\hline
Training $\rightarrow$ Testing & Precision & Accuracy & Recall & F1 score & MCC & AP & Sum \\
\hline
\hline
g-band $\rightarrow$ g-band & ${98.0^{\pm{1.3}}}$ & ${98.8^{\pm{0.3}}}$ &$ {98.2^{\pm{1.2}}}$ &${98.1^{\pm{0.5}}}$ & ${97.2^{\pm{0.8}}}$ & ${99.9^{\pm{0.1}}}$ & ${590.2^{\pm{2.0}}}$
\\
g-band $\rightarrow$ r-band & ${98.5^{\pm{0.7}}}$ & ${97.8^{\pm{0.6}}}$ &$ {94.2^{\pm{2.5}}}$ &${96.3^{\pm{1.1}}}$ & ${94.8^{\pm{1.5}}}$ & ${99.4^{\pm{0.5}}}$ & ${581.0^{\pm{0.8}}}$
\\
r-band $\rightarrow$ r-band & ${97.7^{\pm{1.1}}}$ & ${98.8^{\pm{0.2}}}$ &$ {98.5^{\pm{1.2}}}$ &${98.1^{\pm{0.3}}}$ & ${97.3^{\pm{0.5}}}$ & ${99.7^{\pm{0.4}}}$ & ${590.1^{\pm{1.8}}}$
\\
r-band $\rightarrow$ g-band & ${81.5^{\pm{5.0}}}$ & ${92.9^{\pm{2.1}}}$ &${99.6^{\pm{0.2}}}$ &${89.6^{\pm{2.9}}}$ &${85.3^{\pm{4.0}}}$ & ${99.5^{\pm{0.2}}}$ & ${548.5^{\pm{14.4}}}$\\
\hline
\end{tabular}
\tablefoot{We employ precision, accuracy, recall, F1 score, Matthews correlation coefficient (MCC), and average precision (AP) as performance measurement indicators. The overall performance measurement is obtained by summing the values of these six indicators (shown in the last column). All results listed are SSLS $(R = 3)$.}
\end{center}
\end{table}}

\end{appendix}

\end{document}